\documentclass[Times1COL,AMA]{WileyNJDv5} %STIX1COL,STIX2COL,STIXSMALL

\articletype{BASIC RESEARCH}%

\journal{Journal}
\volume{00}
\copyyear{2025}
\startpage{1}

\definecolor{myblue}{RGB}{0,100,200}
\definecolor{myred}{RGB}{204,102,0}

\usepackage{enumitem}

\usepackage{siunitx}
\DeclareSIUnit\mmHg{mmHg}
\hypersetup{
	colorlinks,
	linkcolor=myred,
	citecolor=myblue,
	urlcolor=myblue
}

\newcommand{\vct}[1]{\boldsymbol{#1}}
\newcommand{\NP}{{\textsf{NP}}}

\newcommand{\omegaNPl}{\omega^{\textsf{NP}\ell}}
\newcommand{\omegaNPv}{\omega^{\textsf{NP}v}}
\newcommand{\omegaNPvhat}{\omega^{\textsf{NP}\hat{v}}}
\newcommand{\omegaNPlD}{\omega^{\textsf{NP}\ell}_D}
\newcommand{\omegaNPvD}{\omega^{\textsf{NP}v}_D}

\newcommand{\soverset}[2]{\overset{\scriptscriptstyle #1}{#2}}

\usepackage[capitalise]{cleveref}
\usepackage{caption}
\usepackage{layouts}

\begin{document}

\title{Computational modelling of cancer nanomedicine: Integrating hyperthermia treatment into a multiphase porous-media tumour model}

\author[1]{Barbara Wirthl*}
\author[2,3]{Paolo Decuzzi}
\author[4,5]{Bernhard A. Schrefler}
\author[1]{Wolfgang A. Wall}

\authormark{WIRTHL \textsc{et al}}

\address[1]{\orgdiv{Institute for Computational Mechanics}, \orgname{Technical University of Munich}, \orgaddress{\city{Garching b. Muenchen}, \country{Germany}}}

\address[2]{\orgdiv{Laboratory of Nanotechnology for Precision Medicine}, \orgname{Italian Institute of Technology}, \orgaddress{\city{Genoa}, \country{Italy}}}

\address[3]{\orgdiv{School of Medicine/Division of Oncology, Center for Clinical Sciences Research}, \orgname{Stanford University}, \orgaddress{\state{CA}, \country{United States}}}

\address[4]{\orgdiv{Department of Civil, Environmental and Architectural Engineering}, \orgname{University of Padua}, \orgaddress{\city{Padua}, \country{Italy}}}

\address[5]{\orgdiv{Institute for Advanced Study}, \orgname{Technical University of Munich}, \orgaddress{\city{Garching b. Muenchen}, \country{Germany}}}

\corres{*Barbara Wirthl, Institute for Computational Mechanics, Technical University of Munich, Boltzmannstrasse 15, 85748 Garching b. Muenchen, Germany. \\ \email{barbara.wirthl@tum.de}}

%\fundingInfo{Text}
%\JELinfo{ejlje}

\abstract[Abstract]{%
    Heat-based cancer treatment, so-called hyperthermia, can be used to destroy tumour cells directly or to make them more susceptible to chemotherapy or radiation therapy.
    To apply heat locally, iron oxide nanoparticles are injected into the bloodstream and accumulate at the tumour site, where they generate heat when exposed to an alternating magnetic field.
    However, the temperature must be precisely controlled to achieve therapeutic benefits while avoiding damage to healthy tissue.
    We therefore present a computational model for nanoparticle-mediated hyperthermia treatment fully integrated into a multiphase porous-media model of the tumour and its microenvironment.
    We study how the temperature depends on the amount of nanoparticles accumulated in the tumour area and the specific absorption rate of the nanoparticles.
    Our results show that host tissue surrounding the tumour is also exposed to considerable doses of heat due to the high thermal conductivity of the tissue, which may cause pain or even unnecessary irreversible damage.
    Further, we include a lumped and a discrete model for the cooling effect of blood perfusion.
    Using a discrete model of a realistic microvasculature reveals that the small capillaries do not have a significant cooling effect during hyperthermia treatment and that the commonly used lumped model based on Pennes' bioheat equation overestimates the effect:
    within the specific conditions analysed, the difference between lumped and discrete approaches is approximatively \SI{0.75}{\celsius}, which could influence the therapeutic intervention outcome.
    Such a comprehensive computational model, as presented here, can provide insights into the optimal treatment parameters for nanoparticle-mediated hyperthermia and can be used to design more efficient treatment strategies.
}

\jnlcitation{}

\maketitle

\renewcommand\thefootnote{}
\footnotetext{\textbf{Abbreviations:} SAR, specific absorption rate; ANC, assemblies of iron oxide nanocubes.}

\renewcommand\thefootnote{\fnsymbol{footnote}}
\setcounter{footnote}{1}

\section{Introduction}\label{sec1}

Hyperthermia therapy is the use of heat to treat cancer, either by destroying tumour cells directly or by making them more susceptible to other treatments, such as radiation therapy or chemotherapy. \cite{Chatterjee2011}
A temperature above $\SI{50}{\celsius}$ causes irreparable coagulation of proteins and other biological molecules and can therefore be used to ablate tumour cells. \cite{Lu2009}
In contrast, a milder rise in temperature in the range of \SIrange{39}{44}{\celsius} shows fewer negative side effects in healthy cells but still has therapeutic benefits \cite{Kaur2016}:
several studies \cite{Jones2005,Issels2010,Herman2013} demonstrated that mild hyperthermia makes tumour cells more susceptible to both radiation therapy and chemotherapy.
In the case of radiation therapy, hyperthermia targets hypoxic cells in the tumour core, which are most sensitive to the cytotoxic effects of heat but resistant to radiation due to the lack of oxygen. \cite{Refaat2015}
In the case of chemotherapy, hyperthermia increases perfusion and thus the delivery of chemotoxic agents, and it also destabilises tumour cells, making them more susceptible to chemotherapy. \cite{Quinto2015}
To achieve these benefits and avoid damage to healthy tissue, heating should be uniform, localised at the tumour site, and must be precisely controlled.

One approach to achieve localised heating is nanoparticle-mediated hyperthermia, particularly using iron oxide nanoparticles. \cite{Palzer2021,Vassallo2023}
The nanoparticles are injected into the bloodstream and accumulate at the tumour site, either passively due to the enhanced permeability and retention (EPR) effect or actively by functionalisation with tumour-targeting ligands or an external magnetic field. \cite{Wilhelm2016}
When exposed to an alternating magnetic field, the nanoparticles generate heat, which is transferred to the surrounding tissue.
Iron, being a ferromagnetic material, is characterised by a high magnetic susceptibility, and when fabricated on the nanoscale ($<\SI{30}{\nano\meter}$), the nanoparticle is composed of only a single magnetic domain.
Without an applied magnetic field, the magnetic moment of the nanoparticles fluctuates randomly and rapidly, and the nanoparticles appear paramagnetic.
In contrast, in the presence of a magnetic field, the nanoparticles show the high magnetic susceptibility of ferromagnetic materials, making them superparamagnetic. \cite{Kaur2016}
An alternating magnetic field can be tuned to resonantly excite superparamagnetic nanoparticles, thereby generating heat via three mechanisms:
Néel relaxation (the rotation of the magnetic moment), Brownian relaxation (the physical rotation of the particle), and hysteresis losses. \cite{Nuzhina2019}

Achieving optimal heating of the tumour site with nanoparticles requires a comprehensive understanding and reliable prediction of two main processes:
the transport of nanoparticles and the subsequent heat generation and transfer.
Advanced computational models can be a powerful tool for studying these processes and optimising treatment parameters.
However, the current state-of-the-art in computational models for nanoparticle-mediated hyperthermia is limited.
Concerning the first process, the transport of nanoparticles, Stillman et al. \cite{Stillman2020} reviewed computational models, ranging from agent-based models to continuum models, that consider transport across various barriers.
But these models do not consider hyperthermia treatment.
Concerning the second process, the heat generation and transfer, multiple reviews \cite{Kaddi2013,Andreozzi2019,Suleman2021} summarised computational models of hyperthermia treatment.
Harry Pennes developed the first heat transfer model \cite{Pennes1948} based on his temperature measurements in the human forearm.
His model, known as Pennes' bioheat equation, includes a lumped term for the cooling effect of blood perfusion.
Because of its mathematical simplicity and pragmatic results and despite its shortcomings, Pennes' bioheat equation has been a standard model for temperature distributions in living tissues and is still widely used. \cite{Nelson1998,Becker2015}
A combined experimental and computational study \cite{Cervadoro2013} used it to investigate the hyperthermic performance of different commercially available superparamagnetic iron oxide nanoparticles.
Extensions of Pennes' bioheat equation include a detailed description of the physics of heat generation \cite{Liangruksa2011} and a model of the cell-nanoparticle interactions and tissue damage. \cite{Huang2010}
Another approach are local thermal equilibrium and local thermal non-equilibrium formulations based on the theory of porous media, as reviewed by Andreozzi et al. \cite{Andreozzi2019}
However, these models focus on heat transfer and lack the transport of nanoparticles in interaction with the tumour microenvironment, which is crucial when studying the effects of nanoparticle-mediated hyperthermia.
Additionally, the cooling effect of blood perfusion is still poorly understood, and the accuracy of the lumped term based on Pennes' bioheat equation is unclear.
Nabil et al. \cite{Nabil2015,Nabil2016} overcame many of the limitations discussed above with a model that couples nanoparticle transport by capillary flow and interstitial filtration with heat transfer and microvascular configurations based on physiological data.
Nevertheless, their model lacks characteristic transport features of the tumour microenvironment, in particular blood vessel collapse, which has a significant impact on nanoparticle accumulation.

To address these shortcomings, we present a computational model for nanoparticle-mediated hyperthermia treatment fully integrated into a multiphase porous-media model of the tumour and its microenvironment.
Our model includes nanoparticle transport in the tumour microenvironment, heat generation by the nanoparticles and heat transfer in the tissue.
In particular, we compare two different models for the cooling effect of blood perfusion: a lumped model based on Pennes' bioheat equation and a discrete model resolving the microvascular network.

\section{Methods}\label{sec2}

In the following, we first give a concise overview of the multiphase porous-media model of the tumour and its microenvironment in Section~\ref{sec:MultiphaseModel} and of the transport of nanoparticles in Section~\ref{sec:TransportNanoparticles}.
Building on these foundations, we introduce the model for nanoparticle-mediated hyperthermia in Section~\ref{sec:HeatEquation}.

\subsection{Multiphase porous-media model of the tumour microenvironment}\label{sec:MultiphaseModel}

The tumour microenvironment is a complex system with various interacting components, such as tumour cells, host cells, the extracellular matrix (ECM), the interstitial fluid (IF), the vasculature and additional subcomponents like oxygen.
To capture the interactions between these components, we use a multiphase porous-media model.
In this contribution, we employ our previously developed model \cite{Sciume2013, Kremheller2018, Kremheller2019} to generate a physically plausible initial condition for the tumour and its microenvironment.
The model has previously been presented, analysed and validated in various forms, including features such as a deformable ECM,\cite{Sciume2014a} invasion of host tissue,\cite{Sciume2014b} and different approaches to model the vasculature and angiogenesis. \cite{Kremheller2018, Kremheller2019}
In the following, we give a brief overview of the model to provide the background for the subsequent modelling of nanoparticle-mediated hyperthermia and refer to the original publications for details on the modelling approach and the governing equations.

The ECM is a mesh-like structure with voids where the cells are attached or migrate and where the fluid flows.
In our multiphase porous-media model, we consider the ECM as the solid phase.
The voids in the ECM constitute the pore space, and the ratio of the volume of the pore space to the total volume is given by the porosity $\varepsilon$.
The IF is modelled as a fluid phase in the pore space.
The cells (tumour and host cells) are also modelled as highly viscous fluids (rather than solids)---similar to most tumour-growth models. \cite{Sciume2013b}
The fluid phases together completely fill, flow in and share the pore space of the ECM.
The fraction occupied by each fluid phase is the saturation $S^\alpha$, defined as
\begin{equation}
    S^\alpha = \frac{\varepsilon^\alpha}{\varepsilon}, \quad \alpha = t,h,\ell
\end{equation}
where $\varepsilon^\alpha$ is the volume fraction of the fluid phase $\alpha$, and the superscripts $t$, $h$, and $\ell$ denote the tumour cell phase, the host cell phase and the IF phase, respectively.
We assume the porous medium to be saturated, i.e., $S^t + S^h + S^\ell = 1$.
All phases can transport chemical subcomponents (so-called species), e.g., oxygen, which are described by the mass fraction $\omega^{i\alpha}$ for a species $i$ in phase $\alpha$.

\begin{figure}[tbp]
    \centering
    \includegraphics[width=\columnwidth]{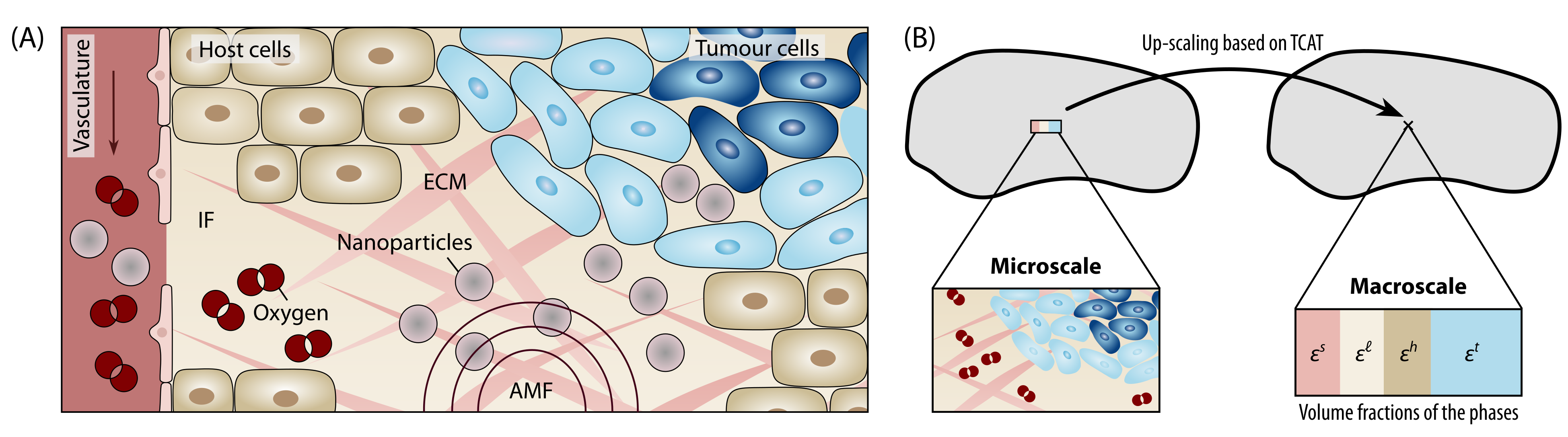}
    \caption{%
        (A)~Porous medium with the pore space of the extracellular matrix (ECM) occupied by the tumour cells, host cells, and the interstitial fluid (IF), and the vasculature as an additional porous network.
        (B)~At the microscale, the different phases can be distinguished (left), while at the macroscale, the phases are described by their volume fractions $\varepsilon^\alpha$ (right).
        Up-scaling based on the thermodynamically constrained averaging theory (TCAT) bridges the gap between the two scales.
    }
    \label{Fig:PorousMedium}
\end{figure}

The ECM, the tumour cells, the host cells and the IF together form the porous medium.
All phases, including their interfaces, can be distinguished at the microscale (see Figure~\ref{Fig:PorousMedium}A).
However, the exact geometry of the ECM is very complex and also not of interest;
neither are we interested in the individual cells.
Our quantity of interest is the tumour as a whole, and we therefore describe it at a larger scale, the macroscale.
At this scale, the different phases are modelled in an averaged sense and characterised by their volume fractions $\varepsilon^\alpha$ at a specific point (see Figure~\ref{Fig:PorousMedium}B).
To bridge the gap between the microscale and the macroscale, we use
the thermodynamically constrained averaging theory (TCAT) \cite{Gray2014} to derive the macroscale equations from the microscale equations while retaining a rigorous connection between the two scales. \cite{Miller2022}

The governing equation for the solid phase is the momentum balance equation.
For the fluid phases, we describe the convective flow with Darcy's law, which allows us to condense the momentum balance equation into the mass balance equation, resulting in a single governing equation.
Species transport is also described by a mass balance equation, including diffusion and advection based on the flow of the fluid phase.
The fluid and species equations are coupled with mass transfer terms that describe, for example, the growth of the tumour or the consumption of oxygen by the proliferating cells. \cite{Sciume2013,Kremheller2018}

The vascular system is another essential part of the tumour microenvironment, as it provides the tumour with oxygen and nutrients and potentially serves as a route for drug delivery.
We include two different models for the vasculature: a homogenised model and a discrete model.
In the homogenised model, we describe the vasculature as an additional porous network in the ECM, resulting in a double-porosity formulation with two separate porous networks.
The first network is the pore space between the ECM fibres with the tumour cells, host cells, and the IF as described above;
the second network is the vasculature, with blood flow and species transport adjacent to the pores of the ECM. \cite{Kremheller2018}
Blood flow is described by the mass balance equation, including convective flow modelled by Darcy's law, similar to the fluid phases.
In the discrete model, the vasculature is modelled as a one-dimensional network of cylindrical pipes, which are embedded in the surrounding porous medium.
Blood flow in the vasculature is described by the Hagen--Poiseuille flow in cylindrical pipes, species transport by a 1D diffusion-advection equation, and mass transfer concentrated as a Dirac measure $\delta_\Lambda$ along the centreline of the pipes. \cite{Kremheller2019}
Here, we either use the homogenised or the discrete model for the vasculature.
A hybrid approach, where the larger vessels are modelled with the discrete model and the smaller vessels with the homogenised model, is also possible. \cite{Kremheller2019,Kremheller2021}

In sum, our multiphase porous-media model is a comprehensive model that captures the physical properties of the different components of the tumour microenvironment, their interactions, and the transport processes.
This provides a physiologically and physically plausible description of the tumour and its microenvironment, which we use as the basis for the subsequent modelling of nanoparticle transport and hyperthermia treatment.

\subsection{Nanoparticle transport}\label{sec:TransportNanoparticles}

On the way from the vasculature to the tumour, nanoparticles experience different transport processes, including extravasation, diffusion in the IF, and advection with the IF flow, which a model for nanoparticle transport must capture.
The model we use has been developed in previous works\cite{Wirthl2020,Wirthl2024,Wirthl2024a}, and we briefly summarise it here to provide the necessary background.

We use a continuum approach to model the transport of the magnetic nanoparticles, employing a diffusion-advection equation directly at the macroscale, because we are not interested in the fate of the individual particles.
The nanoparticles are injected into the bloodstream and transported to the tumour site via the vasculature.
They subsequently extravasate into the IF and travel towards the tumour cells by diffusion and advection with the IF flow.
These different transport processes are described by the mass balance equation of nanoparticles with mass fraction $\omegaNPl$ in the IF given as
\begin{equation}
    \rho^\ell \varepsilon S^\ell \frac{\partial \omegaNPl}{\partial t}
    \bigg\arrowvert_{\vct{X}}
    -\,
    \rho^\ell \frac{\vct{k}^\ell}{\mu^\ell} \vct{\nabla} p^\ell \cdot \vct{\nabla}\omegaNPl
    -
    \vct{\nabla} \cdot \left(
    \rho^\ell\varepsilon S^\ell D^{\NP\ell} \vct{\nabla} \omegaNPl
    \right)
    =
    \sum\limits_{\kappa \in \mathcal{J}_{c\ell}} \soverset{\NP\kappa \rightarrow \NP\ell}{M}
    -
    \omegaNPl \sum\limits_{\kappa \in \mathcal{J}_{c\ell}} \soverset{\kappa  \rightarrow \ell}{M}
    +
    \delta_\Lambda
    \left(
    \soverset{\NP\hat{v} \rightarrow \NP\ell}{M}
    -
    \omegaNPl \soverset{\hat{v} \rightarrow \ell}{M}
    \right)
    \label{Eq:MassBalanceSpeciesTransport}
\end{equation}
similar to the other species.
Herein, $p^\ell$ denotes the pressure and $\rho^\ell$ the density.
The diffusivity of nanoparticles is given by $D^{\NP\ell}$, and advection with the IF flow is again described by Darcy's law, with $\vct{k}^\ell$ denoting the permeability of the ECM with respect to IF and $\mu^\ell$ the viscosity of the IF.
The terms on the right-hand side of Equation (\ref{Eq:MassBalanceSpeciesTransport}) describe the mass transfer of nanoparticles to and from the IF, where the superscript $v$ denotes the homogenised vasculature and $\hat{v}$ the discrete vasculature.
The last two terms describe the mass transfer from the discretely modelled vasculature to the IF and are therefore scaled with the Dirac measure $\delta_\Lambda$ along the centreline of the vessels.

For the mass transfer, we include extravasation via the interendothelial and the transendothelial pathway and lymphatic drainage, as given by
\begin{equation}
    \sum\limits_{\kappa \in \mathcal{J}_{c\ell}} \soverset{\NP\kappa \rightarrow \NP\ell}{M}
    =
    \soverset{\NP v \rightarrow \NP\ell}{M_{\textsf{inter}}}
    \; + \;
    \soverset{\NP v \rightarrow \NP\ell}{M_{\textsf{trans}}}
    \; - \;
    \soverset{\NP\ell \rightarrow \NP{ly}}{M_{\textsf{drain}}}.
\end{equation}
The extravasation of nanoparticles from the vasculature into the IF occurs through two different pathways:
the interendothelial and the transendothelial pathway. \cite{Wilhelm2016,Moghimi2018}
The interendothelial pathway is a convective process, meaning the transvascular fluid flow drags the nanoparticles \cite{Jain1987}:
the tumour vasculature is leaky and hyperpermeable due to poorly aligned endothelial cells, which results in gaps between adjacent cells through which fluid leaks. \cite{Jain2010}
The transendothelial pathway describes the diffusion of nanoparticles through the vessel wall, for example, through interconnected cytoplasmic vesicles and vacuoles, driven by the concentration gradient of nanoparticles across the vessel wall. \cite{Wilhelm2016}
In the homogenised model, the mass transfer of nanoparticles from the vasculature into the IF is given by
\begin{subequations}
    \label{Eq:CrossingBloodVesselWallHomogenised}
    \begin{align}
        \soverset{\NP v \rightarrow \NP\ell}{M_{\textsf{inter}}}
         & =
        \rho^v \varepsilon^v L_p^v \frac{S}{V}
        \left[
            p^v - p^\ell - \sigma \left( \pi^v - \pi^\ell \right)
            \right] \tfrac{\omegaNPv + \omegaNPl}{2}
        \\
        \soverset{\NP v \rightarrow \NP\ell}{M_{\textsf{trans}}}
         & =
        \rho^v \varepsilon^v P^{v} \frac{S}{V} \left\langle \omegaNPv - \omegaNPl \right\rangle _+,
    \end{align}
\end{subequations}
with the hydraulic conductivity $L_p^v$, the surface-to-volume ratio $S/V$, the oncotic pressure difference $\sigma (\pi^v - \pi^\ell)$, and the permeability $P^{v}$.
In the discrete model, the mass transfer is given by
\begin{subequations}
    \label{Eq:CrossingBloodVesselWallDiscrete}
    \begin{align}
        \soverset{\NP \hat{v} \rightarrow \NP\ell}{M_{\textsf{inter}}}
         & =
        \rho^v \, 2 \pi R \, L_p^v
        \left[
            p^{\hat{v}} - p^\ell - \sigma \left( \pi^v - \pi^\ell \right)
            \right] \tfrac{\omegaNPvhat + \omegaNPl}{2}
        \\
        \soverset{\NP \hat{v} \rightarrow \NP\ell}{M_{\textsf{trans}}}
         & =
        \rho^v \, 2 \pi R \, P^v \left\langle \omegaNPvhat - \omegaNPl \right\rangle _+,
    \end{align}
\end{subequations}
with the radius $R$ of the blood vessel.
Note that because mass transfer occurs across the entire vessel wall, the mass transfer in the discrete model is scaled by the circumference of the vessel.

In addition, lymphatic vessels contribute to mass transfer from the IF.
While the lymphatic vessels absorb extravasated fluid and molecules in normal tissues, they are impaired in tumours, resulting in inefficient drainage. \cite{Carmeliet2000,Jain2002}
The uptake of nanoparticles by the lymphatic system is described by
\begin{equation}
    \overset{\NP\ell \rightarrow \NP{ly}}{M_{\textsf{drain}}}
    =
    \rho^\ell \left(L_p \frac{S}{V}\right)^{ly} \left\langle p^\ell - p^{ly} \right\rangle _+ \left\langle 1 - \frac{p^t}{p^{ly}_{\textsf{coll}}} \right\rangle _+ \omegaNPl
\end{equation}
with the lymphatic filtration coefficient $\left(L_p \frac{S}{V}\right)^{ly}$ and the lymphatic pressure $p^{ly}$ which we assume to be zero.
Above the collapsing pressure $p^{ly}_{\textsf{coll}}$, lymphatic drainage is impaired and no fluid or particles are taken up by the lymphatic system.

In sum, our model of nanoparticle transport includes all major transport processes and barriers that nanoparticles encounter to reach the tumour site.
Since the amount of heat generated by excitation of the nanoparticles depends on where and how many nanoparticles accumulate in the tumour area, a physically and physiologically appropriate model of nanoparticle transport is essential to predict the temperature during nanoparticle-mediated hyperthermia treatment.

\subsection{Heat transfer}\label{sec:HeatEquation}

In our model of heat generation and transfer during hyperthermia treatment, we assume that all phases are locally in a state of thermodynamic equilibrium, and hence the temperature $T$ of all phases is equal, i.e.,
\begin{equation}
    T^\gamma = T,
    \quad \forall \gamma,
    \quad
    \gamma \in \left\{ s, t, h, \ell, v \right\},
\end{equation}
where the index $\gamma$ denotes the phases.
In addition to the indices for the fluid phases and the vasculature, the index $s$ denotes the solid phase.
The balance equation for the temperature is the energy balance given as enthalpy balance \cite{Pesavento2016}
\begin{equation}
    c_{p}^{\gamma}\frac{\partial \left( \rho^\gamma \varepsilon^\gamma T \right)}{\partial t}
    \bigg\arrowvert_{\vct{x}}
    +
    c_{p}^{\gamma} \vct{\nabla} \cdot \left( \rho^\gamma \varepsilon^\gamma \, T \vct{v}^\gamma \right)
    -
    \vct{\nabla} \cdot \left( \kappa^\gamma \varepsilon^\gamma \vct{\nabla} T \right)
    =
    \varepsilon^\gamma \left(Q_p - Q_{bl}\right),
\end{equation}
with the specific heat capacity $c_{p}^{\gamma}$, the thermal conductivity $\kappa^\gamma$, and the velocity $\vct{v}^\gamma$ of phase $\gamma$.
We include a heat source $Q_p$ due to heat generated by the nanoparticles and a heat sink $Q_{bl}$ due to blood perfusion, which we further detail below.
The time derivative is evaluated at a spatial coordinate $\vct{x}$ as opposed to a material coordinate $\vct{X}$.
We neglect viscous dissipation, mechanical work, density variation, and kinetic energy.
Now, we apply the product rule to the time derivative and the convective term, transform the spatial time derivative to a material time derivative and apply the mass balance to obtain
\begin{subequations}
    \label{Eq:HeatEquationTransformed}
    \begin{align}
        c_{p}^{\,s} \rho^s \varepsilon^s \frac{\partial T}{\partial t}
        \bigg\arrowvert_{\vct{X}}
        -
        \vct{\nabla} \cdot \left(\kappa^s \varepsilon^s \vct\nabla T \right)
         & =
        \varepsilon^s \left(Q_p - Q_{bl}\right)
        -
        c_{p}^{\,s} T \sum\limits_{\kappa \in \mathcal{J}_{cs}} \overset{\kappa \rightarrow s}{M},
        \\
        c_{p}^{\tau} \rho^\tau \varepsilon^\tau \frac{\partial T}{\partial t}
        \bigg\arrowvert_{\vct{X}}
        +
        c_{p}^{\tau} \rho^\tau \varepsilon^\tau \left( \vct{v}^\tau - \vct{v}^{s} \right) \cdot \vct{\nabla} T
        -
        \vct{\nabla} \cdot \left( \kappa^\tau \varepsilon^\tau \vct{\nabla} T \right)
         & =
        \varepsilon^\tau \left(Q_p - Q_{bl}\right)
        -
        c_{p}^{\tau} T \hspace{-1mm} \sum\limits_{\kappa \in \mathcal{J}_{c\tau}}\hspace{-1mm} \overset{\kappa \rightarrow \tau}{M},
        \quad
        \tau \in \left\{ t, h, \ell, v \right\}.
    \end{align}
\end{subequations}
A step-wise derivation is given in Appendix~\ref{App:SuppEquations}.
Note that the convective term cancels out for the solid phase.
Finally, we sum Equations~(\ref{Eq:HeatEquationTransformed}) over all phases $\gamma$ and get
\begin{equation}
    \left( c_p \rho \right)_{\textsf{eff}} \frac{\partial T}{\partial t}
    \bigg\arrowvert_{\vct{X}}
    +
    \sum \limits_{\tau} \left[ c_{p}^{\tau} \rho^{\tau}\varepsilon^{\tau} \left( \vct{v}^{\tau} - \vct{v}^s \right) \cdot \vct{\nabla} T \right]
    -
    \vct{\nabla} \cdot \left( \kappa_{\mathrm{eff}} \vct{\nabla}T\right) \\
    =
    Q_p
    -
    Q_{bl}
    -
    \sum\limits_{\gamma}  \Bigl[ \, c_{p}^{\gamma} T \sum\limits_{\kappa \in \mathcal{J}_{c\gamma}} \overset{\kappa \rightarrow \gamma}{M} \, \Bigr].
\end{equation}
The last term on the right-hand side cancels out because the sum of mass transfer terms over all phases is zero.
Further, the effective heat capacity is given as
\begin{equation}
    \left( c_p \rho \right)_{\textsf{eff}}
    =
    \sum\limits_{\gamma} c_{p}^{\gamma} \rho^\gamma \varepsilon^\gamma
\end{equation}
and the effective thermal conductivity as
\begin{equation}
    \kappa_{\textsf{eff}}
    =
    \sum\limits_{\gamma} \kappa^{\gamma} \varepsilon^\gamma.
\end{equation}

The heat source term $Q_p$ models the heat that is generated by the nanoparticles, in our case iron oxide nanoparticles exposed to an alternating magnetic field.
The specific absorption rate ($SAR$) quantifies the efficacy of nanoparticles in generating heat when exposed to the alternating magnetic field and depends on the frequency, the magnetic field strength, and other parameters such as the nanoparticle diameter. \cite{Fortin2007}
We assume that the amount of heat that is generated is directly proportional to the mass fraction of nanoparticles.
Following studies in the literature, \cite{Cervadoro2013,Nabil2015} we model the heat source term as
\begin{equation}
    Q_p
    =
    \left(
    \rho^v \varepsilon^v \omegaNPv
    +
    \rho^\ell \varepsilon S^\ell \omegaNPl
    +
    \rho^v \pi R^2  \omegaNPvhat \delta_\Lambda
    \right) {SAR},
\end{equation}
where the contribution of nanoparticles in the discrete vasculature $\omegaNPvhat$ is only non-zero in the discrete model of the vasculature and is scaled with the Dirac measure $\delta_\Lambda$.
Note that the contribution of nanoparticles in the discrete vasculature is scaled with the cross-sectional area of the blood vessel.
For the heat sink due to blood perfusion, we again consider two different forms: one for the homogenised and one for the discrete model of the vasculature.
In the homogenised case, we adopt Pennes' bioheat equation \cite{Pennes1948} and include the cooling effect of blood perfusion as a spatially averaged lumped heat sink term in the form of
\begin{equation}
    Q_{bl}
    =
    \rho^v c_p^v w \left( T - T_{b} \right) \vphantom{\left( \varepsilon^v \omegaNPv + \varepsilon S^l \omegaNPl \right)},
\end{equation}
where $w$ is the blood perfusion rate and $T_{b}$ the body temperature.
Considering that during local hyperthermia treatment heat generation is localised at the tumour site, it is reasonable to assume body and blood temperature homeostasis.
Hence, we assume the body and the blood temperature to be constant at normal body temperature $T_{b} = \SI{37}{\degreeCelsius}$.
In the discrete case, we consider the heat sink due to blood perfusion similar to Nabil et al. \cite{Nabil2015,Nabil2016} in the form of
\begin{equation}
    Q_{bl}
    =
    2 \pi R \beta_T \left( T - T_{b} \right) \delta_\Lambda,
    \label{Eq:HeatSinkDiscrete}
\end{equation}
with the heat exchange coefficient $\beta_T$.
Again, note that the heat sink term in the discrete model is scaled with the circumference of the blood vessel, similar to the mass transfer of nanoparticles.
This results in a heat transfer formulation that is similar to the local thermal non-equilibrium formulations in the literature. \cite{Yuan2008,Andreozzi2019}

As boundary condition for the energy balance, we apply a Robin-type boundary condition at the outer boundary of the domain, which accounts for heat exchange with the surrounding tissue and is given as \cite{Nabil2015, Nabil2016}
\begin{equation}
    \label{Eq:RobinBoundaryCondition}
    -\kappa_{\textsf{eff}} \; \vct{n}
    =
    \beta_T \left( T - T_{b} \right),
\end{equation}
with the outer unit normal vector $\vct{n}$ and the heat exchange coefficient $\beta_T$ accounting for heat flux to the surrounding tissue, which is assumed to be at body temperature $T_{b}$.

Altogether, our model for nanoparticle-mediated hyperthermia treatment considers heat transfer due to diffusion and convection in the tissue, heat generation by the nanoparticles, and the cooling effect of blood perfusion.

\subsection{Computational solver}

We solve the coupled system of equations using the finite element method (FEM):
we apply the standard Galerkin method to obtain the weak form of the governing equations, i.e., we multiply the equations by test functions, integrate over the domain, and apply Gauss' theorem to terms containing a second spatial derivative to decrease differentiability requirements of the solution function space.
For discretisation in time, we use the backward Euler method with a time step size $\Delta t$ and initial conditions specified at $t = 0$.
For discretisation in space, we use quadrilateral elements with bilinear shape functions.
Further, we employ a monolithic approach to solve the coupled system of equations. \cite{Kremheller2018}
This results in a potentially large system of nonlinear equations, which we solve using a single Newton--Raphson loop per time step.
As a linear solver, we use a generalised minimal residual (GMRES) iterative solver combined with an algebraic multigrid (AMG) preconditioner. \cite{Verdugo2016,Fang2019}
More details on the FEM can be found in standard textbooks. \cite{Wriggers2008a,Zienkiewicz2013}
The implementation of the model and the solver are available as part of our open-source project 4C. \cite{4C2025}

\section{Results and discussion}\label{sec3}

In the following, we study nanoparticle-mediated hyperthermia treatment in three different scenarios.
First, in Section~\ref{sec:IdealisedSphericalTumour}, we analyse an idealised spherical tumour and conduct a parameter study to understand the influence of different parameters on the temperature increase during treatment, in particular of the lumped heat sink term based on Pennes' bioheat equation.
Second, in Section~\ref{sec:TumourDiscreteVasculature}, we investigate the cooling effect of a microvascular network using the discrete model of the vasculature.
Finally, in Section~\ref{sec:MouseModel}, we study the influence of clustering of nanoparticles in an \textit{in vivo} mouse model.

\subsection{Idealised spherical tumour with lumped heat sink term}
\label{sec:IdealisedSphericalTumour}

We first analyse nanoparticle-mediated hyperthermia treatment of an idealised spherical tumour growing in a vascularised host tissue.
In this idealised scenario, we employ the homogenised model for the vasculature.
We study the temperature increase depending on the mass fraction of injected nanoparticles, the specific absorption rate ($SAR$), and the blood perfusion rate.
In particular, we investigate the influence of the lumped heat sink term due to blood perfusion as commonly employed in Pennes' bioheat equation.

To generate a physically and physiologically plausible tumour microenvironment, we employ our previously developed tumour-growth model, as summarised in Section~\ref{sec:MultiphaseModel}.
This example of an idealised spherical tumour in its microenvironment is based on our previous publication, \cite{Wirthl2020} where the setup including all parameters, boundary and initial conditions is described in detail.
We analyse a domain of $\SI{1}{\milli\meter}\times \SI{1}{\milli\meter}$ where, due to the symmetry of the problem, only one quarter is actually simulated ($\SI{0.5}{\milli\meter}\times \SI{0.5}{\milli\meter}$).
The domain is discretised with $120 \times 120$ elements, and the structure, fluid and species transport meshes are conforming.

\begin{figure}[btp]
    \centering
    \includegraphics[width=0.95\textwidth]{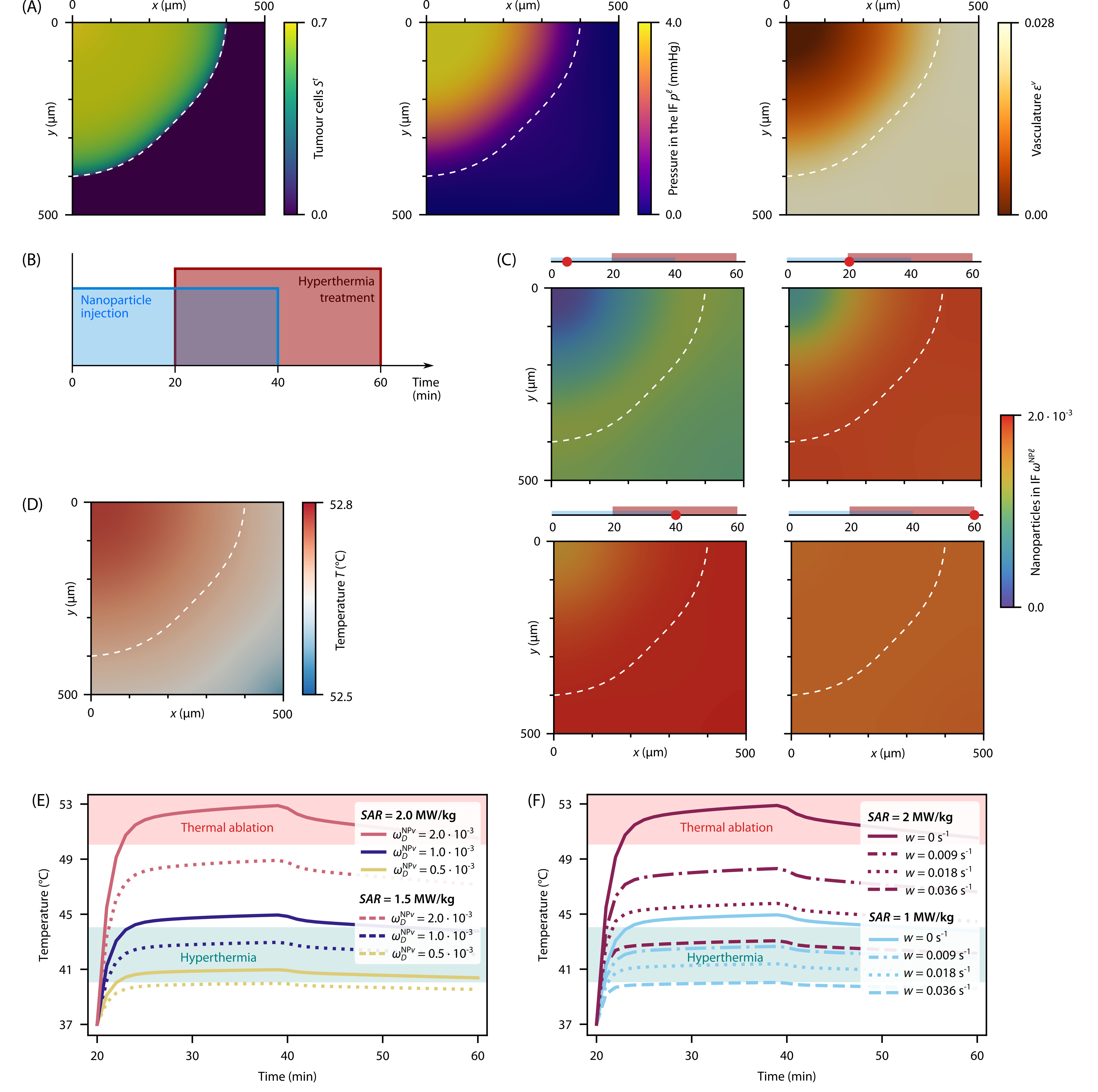}
    \caption{Idealised spherical tumour with lumped heat sink term.
        (A)~Characteristic features of a solid tumour described by the saturation of tumour cells $S^t$, the pressure in the interstitial fluid (IF) $p^\ell$, and the volume fraction of the vasculature $\varepsilon^v$.
        The white dashed line indicates the tumour boundary in all plots.
        (B)~Treatment protocol.
        (C) Mass fraction of nanoparticles in the IF $\omegaNPl$ after \SI{5}{\minute}, \SI{20}{\minute}, \SI{40}{\minute}, and \SI{60}{\minute}.
        (D)~Temperature field after \SI{40}{\minute}.
        (E)~Temperature curves for different values of injected nanoparticles $\omegaNPvD$ and specific absorption rates ($SAR$).
        (F)~Temperature curves for different specific absorption rates ($SAR$) and blood perfusion rates $w$.
    }
    \label{fig:IdealisedSphericalTumour}
\end{figure}

In its grown state, which is shown in Figure~\ref{fig:IdealisedSphericalTumour}A, the tumour has a radius of $\SI{400}{\micro\meter}$ and exhibits characteristic features known from solid tumours.
The interstitial pressure in the tumour is elevated, reaching a maximum of $p^\ell = \SI{4}{\mmHg}$, which is in the range of values reported in the literature. \cite{Boucher1990, Jain2007, Dewhirst2017}
This elevated interstitial pressure generates an outward flow of interstitial fluid, which is an obstacle in tumour treatment as it hinders the transport of drugs and nanoparticles. \cite{Heldin2004}
Further, the growing tumour pushes against its surrounding microenvironment, thereby collapsing blood vessels---a hallmark shared by all solid tumours \cite{Stylianopoulos2012,Provenzano2012,Chauhan2014}:
while the surrounding host tissue is vascularised with a volume fraction of $\varepsilon^v = 0.028$, the tumour has a non-perfused core.
This lack of perfusion is an additional major challenge as it limits the delivery of drugs to the tumour site.
These features of the tumour microenvironment have a decisive influence on the transport of nanoparticles and thus on the hyperthermia treatment, which we investigate in the following.

To study the temperature increase during nanoparticle-mediated hyperthermia treatment, we consider the following \textit{in silico} treatment protocol based on Nabil et al., \cite{Nabil2015} as shown in Figure~\ref{fig:IdealisedSphericalTumour}B:
we analyse a 60-minute period in which nanoparticles are injected into the vasculature for the first 40 minutes and then exposed to an alternating magnetic field for the interval between 20 and 60 minutes, so that the nanoparticles generate heat.
We assume that the nanoparticles are continuously injected and that the infusion directly affects the blood concentration in the entire systemic circulation. \cite{Nabil2015}
Therefore, we prescribe the mass fraction of nanoparticles in the vasculature $\omegaNPvD = \num{2.0e-3}$, $\num{1.0e-3}$, and $\num{0.5e-3}$ as Dirichlet boundary condition.
These values match the injected concentrations used in the experiment performed by Cervadoro et al. \cite{Cervadoro2013}
Here and in the following, we use the subscript $D$ to denote values applied as Dirichlet boundary conditions.
Concerning the thermal properties of the tissue, we assume the heat capacity and thermal conductivity to be identical for all phases.
The parameters are listed in Table~\ref{Tab:ParametersNanoparticleHyperthermia} and are based on the literature and experimental data.
As boundary condition for the temperature, we apply the Robin-type boundary condition at the outer boundary of the domain, i.e., the bottom and right boundary of the domain, to account for heat exchange with the surrounding tissue.
The initial temperature is the normal body temperature $T_{b} = \SI{37}{\celsius}$.
The time step is set to $\Delta t = \SI{60}{\second}$, and we simulate 60 time steps.
All other parameters for the tumour microenvironment are identical to the ones used to generate the initial condition and can be found in the original publication. \cite{Wirthl2020}

\begin{table}[btp]
    \caption{Parameters for nanoparticles transport and hyperthermia treatment simulations for the idealised spherical tumour (Section~\ref{sec:IdealisedSphericalTumour}) and the tumour with discrete vasculature (Section~\ref{sec:TumourDiscreteVasculature}).}
    \label{Tab:ParametersNanoparticleHyperthermia}
    \centering
    \begin{tabular}{llllll}
        \toprule
        \multicolumn{2}{l}{Parameter}                       & Value                                    & Unit                                                   & Source                                                                                         \\
        \midrule
        \multicolumn{3}{l}{\textbf{Nanoparticle transport}} &                                                                                                                                                                                                    \\
        $D^{\NP\ell}$                                       & Diffusivity of nanoparticles in the IF   & $\num{1.2955e-5}$                                      & \si[per-mode=symbol]{\milli\meter\squared\per\second}          & \citenum{Nabil2015}           \\
        $P^v$                                               & Blood vessel wall permeability           & $\num{2.0e-6}$                                         & \si[per-mode=symbol]{\milli\meter\per\second}                  & \citenum{Nabil2015}           \\
        $L_p^v$                                             & Blood vessel wall hydraulic conductivity & $\num{1.0e-7}$                                         & \si[per-mode=symbol]{\milli\meter\squared\second\per\gram}     & \citenum{Nabil2015}           \\
        $(L_p \; S/V)^{ly}$                                 & Lymphatic filtration coefficient         & $\num{1.04e-06}$                                       & \si[per-mode=symbol]{\per\pascal\per\second}                   & \citenum{Baxter1990}          \\
        \multicolumn{3}{l}{\textbf{Hyperthermia treatment}} &                                                                                                                                                                                                    \\
        $T_{b}$                                             & Body temperature                         & $\num{310.15}$                                         & \si{\kelvin}                                                   & Known                         \\
        $c_p^\gamma \;\forall \gamma$                       & Tissue-specific heat capacity            & $\num{3470}$                                           & \si[per-mode=symbol]{\joule\per\kilo\gram\per\kelvin}          & \citenum{Nabil2015,Huang2010} \\
        $\kappa^\gamma \;\forall \gamma$                    & Tissue thermal conductivity              & $\num{0.51e-3}$                                        & \si[per-mode=symbol]{\watt\per\milli\meter\per\kelvin}         & \citenum{Cervadoro2013}       \\
        $SAR$                                               & Specific absorption rate                 & $\num{1.0}$, $\num{1.5}$, $\num{2.0}$                  & \si[per-mode=symbol]{\mega\watt\per\kilo\gram}                 & \citenum{Nabil2015}           \\
        $w$                                                 & Perfusion rate                           & $\num{0}$, $\num{0.009}$, $\num{0.018}$, $\num{0.036}$ & \si[per-mode=reciprocal]{\per\second}                          & \citenum{Cervadoro2013}       \\
        $\beta_T$                                           & Heat exchange coefficient                & $\num{2e-5}$                                           & \si[per-mode=symbol]{\watt\per\milli\meter\squared\per\kelvin} & \citenum{Nabil2016}           \\
        \bottomrule
    \end{tabular}
\end{table}

Figure~\ref{fig:IdealisedSphericalTumour}C show the resulting mass fraction of nanoparticles in the IF for different time points, as an example for an injected mass fraction of nanoparticles in the vasculature of $\omegaNPvD = \num{2.0e-3}$.
During the \SI{40}{\minute} of nanoparticle injection, more nanoparticles accumulate in the well-perfused area of the domain, i.e., the nanoparticles accumulate close to where they crossed the blood vessel wall.
Fewer nanoparticles reach the non-perfused tumour core.
After the injection phase, the nanoparticles further diffuse in the IF, reaching a homogeneous distribution after \SI{60}{\minute}.
At the same time, the lymphatic drainage removes nanoparticles;
hence, the mass fraction decreases.

Concerning the temperature field, Figure~\ref{fig:IdealisedSphericalTumour}D shows that the temperature is relatively homogeneous with a temperature difference of only $\Delta T = \SI{0.3}{\celsius}$.
Hence, not only the tumour but also the surrounding healthy tissue is exposed to high temperatures over \SI{52}{\celsius}.
This is due to the high thermal conductivity of the tissue and the accumulation of nanoparticles in healthy tissue around the tumour.

In the following parameter study, we investigate the dependence of the temperature increase on the mass fraction of injected nanoparticles $\omegaNPvD$, the specific absorption rate ($SAR$), and the blood perfusion rate $w$.
Figure \ref{fig:IdealisedSphericalTumour}E shows the resulting temperature increase during hyperthermia treatment (\SIrange{20}{60}{\minute}) for the different parameter values.
All curves present the average temperature of the entire domain.
In all cases, the temperature rises steeply at the beginning of the treatment and then asymptotically approaches a steady state, reaching the maximum temperature at \SI{40}{\minute}.
Thereafter, the nanoparticle injection stops, and the temperature starts to decrease.

First, we analyse the temperature increase for the three different mass fractions of injected nanoparticles $\omegaNPvD$ with specific absorption rates of $SAR = \SI{2.0}{\mega\watt\per\kilo\gram}$ and $SAR = \SI{1.5}{\mega\watt\per\kilo\gram}$.
For all these simulations, we use a blood perfusion rate of $w = 0$.
The results are shown in Figure~\ref{fig:IdealisedSphericalTumour}E.
At the highest mass fraction of $\omegaNPvD = \num{2.0e-3}$, the temperature reaches a maximum of $\SI{53}{\celsius}$ at $\SI{40}{\minute}$, which is in the range of thermal ablation.
In contrast, the temperature stays in the range of \SIrange{40}{41}{\celsius} for the lowest mass fraction of injected nanoparticles of $\omegaNPvD = \num{0.5e-3}$.
Next, we compare the results for the two different specific absorption rates $SAR = \SI{2.0}{\mega\watt\per\kilo\gram}$ and $\SI{1.5}{\mega\watt\per\kilo\gram}$, which are of the same order of magnitude as the values presented in the literature. \cite{Nabil2015}
The higher specific absorption rate results in a higher temperature increase, which is suitable for thermal ablation.
The lower specific absorption rate results in a temperature increase mainly suitable for hyperthermia treatment but not for thermal ablation.
This demonstrates the challenge associated with nanoparticle-based hyperthermia treatment:
the temperature highly depends on the amount of nanoparticles accumulating in the tumour and the properties of the nanoparticles, in particular their specific absorption rate.
Both must be precisely controlled to achieve the desired temperature increase and avoid under- or overtreatment.

Finally, we investigate the influence of the blood perfusion rate as a heat sink on the temperature increase.
For two specific absorption rates $SAR = \SI{2.0}{\mega\watt\per\kilo\gram}$ and $\SI{1.0}{\mega\watt\per\kilo\gram}$, we analyse the temperature increase for four different blood perfusion rates $w = \SI[per-mode=reciprocal]{0}{\per\second}$, $\SI[per-mode=reciprocal]{0.009}{\per\second}$, $\SI[per-mode=reciprocal]{0.018}{\per\second}$, and $\SI[per-mode=reciprocal]{0.036}{\per\second}$, based on the literature. \cite{Cervadoro2013}
The results are also shown in Figure \ref{fig:IdealisedSphericalTumour}F.
The blood perfusion based on Pennes' bioheat equation also reduces the temperature increase.
For the higher specific absorption rate (purple lines), the temperature increase is reduced by $\SI{10}{\celsius}$ for a blood perfusion rate of $w = \SI[per-mode=reciprocal]{0.036}{\per\second}$ compared to the case without blood perfusion.
This effect diminishes for the lower specific absorption rate (blue lines) to a reduction by $\SI{5}{\celsius}$.

The investigation of temperature increase in and around the tumour confirms that hyperthermia (\SIrange{39}{44}{\celsius}) and thermal ablation ($>$\SI{50}{\celsius}) both can be reached.
The particles can be produced with different specific absorption rates and with different injection concentrations, which allows fine-tuning to the desired temperature range. \cite{Wilhelm2016}
Here, we describe heat generation by excitation of nanoparticles in a single equation based on the specific absorption rate, similar to Nabil et al. \cite{Nabil2015}
If a detailed description of the physics behind the heating process should prove necessary, the heat source term can be replaced by the model presented by Liangruksa et al. \cite{Liangruksa2011}, which includes physical details, e.g., the Néelian and Brownian mechanisms of relaxation and the amplitude and frequency of the alternating magnetic field.

One major advantage of using nanoparticles for hyperthermia treatment is the fact that it allows heating of the tumour while reducing damage to normal tissue, because the temperature decreases rapidly with increasing distance from the heat source. \cite{Chatterjee2011}
Our results however show that healthy tissue surrounding the tumour is still exposed to considerable doses of heat due to the high thermal conductivity of the tissue.
Such high temperatures in healthy tissue may cause pain or even unnecessary irreversible damage.

We here employ a lumped heat sink term based on Pennes' bioheat equation to model the cooling effect of blood perfusion.
This simple lumped model triggered a controversy in the literature.
One major deficiency is the lumped spatially averaged formulation for heat transfer via blood perfusion, which assumes a uniform perfusion rate without considering the direction of blood flow.
According to Becker and Kuznetsov, \cite{Becker2015} the Pennes' bioheat equation is an approximation equation without a physically consistent theoretical basis.
Wulff \cite{Wulff1974} stated that this description is inconsistent and results in errors of the same order of magnitude as the convective energy transport itself.
An alternative equation, presented by Weinbaum and Jiji, \cite{Weinbaum1985} is based on the fact that small arteries and veins are parallel:
the flow direction is countercurrent and heating and cooling effects are counterbalanced.
More recent models---including our model---consider human tissue as a porous medium. \cite{Khaled2003,Nakayama2008}
While here we assume that all phases are in thermodynamic equilibrium, our model can easily be extended to consider phases with different temperatures.

Given all this criticism, one may ask why the Pennes' bioheat equation is still widely used and accepted for numerical simulation of hyperthermia treatment.
An analysis \cite{Wissler1998} pointed out significant problems with Pennes' procedure to analyse his data, but at the same time shows that a more rigorous examination still yields good agreement with the model.
Finally, a comparison \cite{Hassanpour2014} of Pennes' bioheat model to the counter-current model and to porous media models showed that despite different temperature fields, all three models predict similar heat-affected zones, which is the crucial point for hyperthermia treatment.

In summary, our parameter study shows that the temperature increase highly depends on the mass fraction of nanoparticles accumulating in the tumour area and on the specific absorption rate of the nanoparticles.
To further investigate the validity of the lumped heat sink term in the modelling of hyperthermia treatment of tumours, in the following section we will compare the results discussed in this section to results with the cooling effect of blood perfusion discretely resolved.

\subsection{Tumour with a discrete microvascular network}
\label{sec:TumourDiscreteVasculature}

\begin{figure}[btp]
    \centering
    \includegraphics[width=\textwidth]{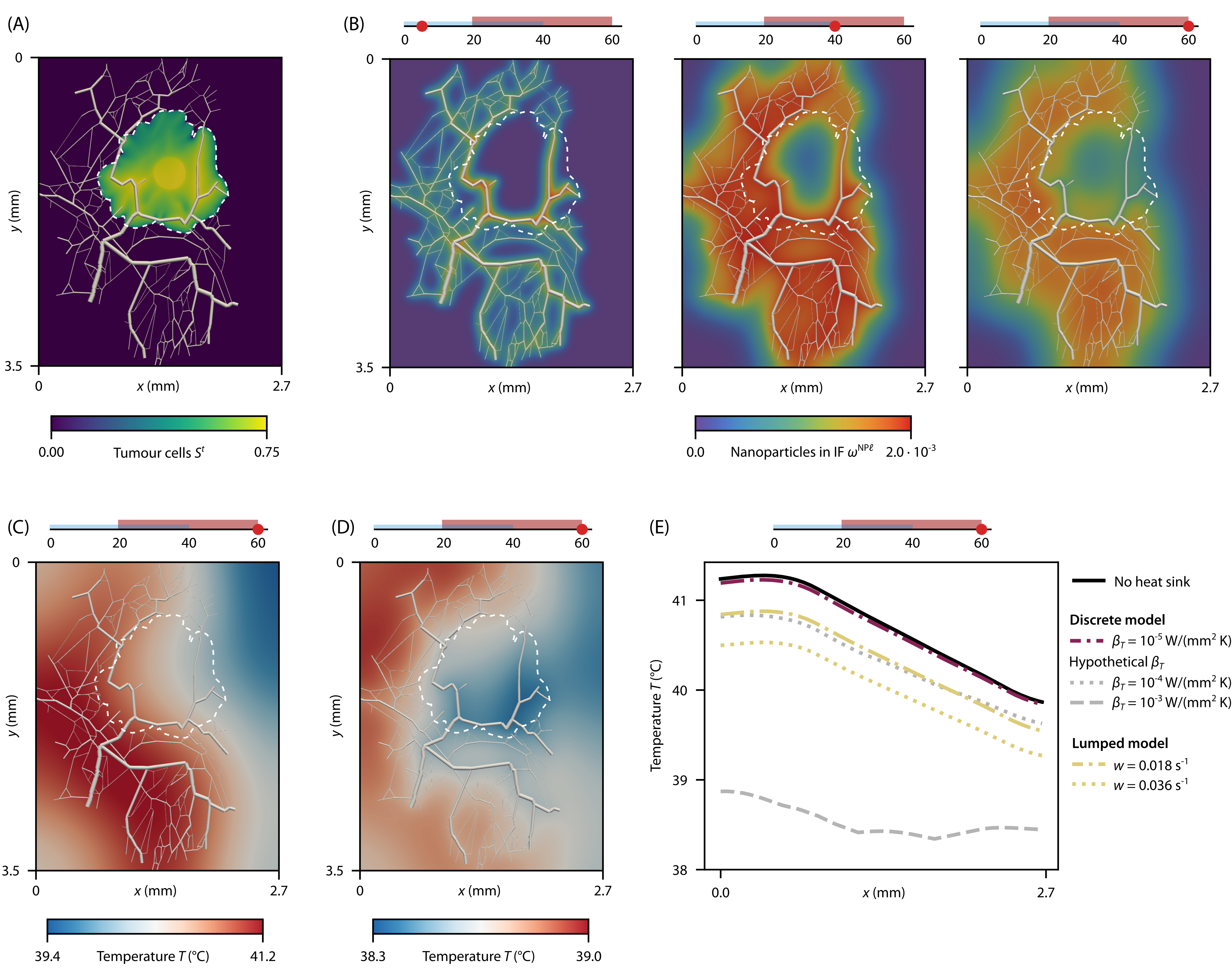}
    \caption{Tumour with a discrete microvascular network.
        (A)~Saturation of tumour cells $S^t$.
        The white dashed line indicates the tumour boundary in all plots.
        (B)~Mass fraction of nanoparticles in the interstitial fluid (IF) $\omegaNPl$ after \SI{5}{\minute}, \SI{40}{\minute}, and \SI{60}{\minute}.
        (C)~Temperature field after \SI{60}{\minute} for a heat exchange coefficient $\beta_T = \SI{2e-5}{\watt\per\milli\meter\squared\per\kelvin}$.
        (D)~Temperature field after \SI{60}{\minute} for a hypothetical heat exchange coefficient $\beta_T = \SI{2e-3}{\watt\per\milli\meter\squared\per\kelvin}$.
        (E)~Comparison of the temperature curves at $y = \SI{1.8}{\milli\meter}$ after \SI{60}{\minute} for the discrete and the lumped model of the cooling effect of blood perfusion, including different values for the heat exchange coefficient $\beta_T$ or the blood perfusion rate $w$ and the case without blood perfusion.
    }
    \label{fig:WithArteries}
\end{figure}

We now investigate nanoparticle transport and the temperature increase during hyperthermia treatment with the discrete model of the vasculature.
We analyse a domain of $\SI{2.7}{\milli\meter}\times \SI{3.5}{\milli\meter}$ with a microvascular network based on \textit{in vivo} data.
The network topology is based on experimental data \cite{Pries1990,Walker-Samuel2017} of a microvascular network of a rat, which is adapted to the size of the domain, i.e., the geometry is scaled down by a factor of two while maintaining the original vessel radii:
the network is fed by one major arteriole, and the vessel radii are in the range of \SIrange{1.6}{30}{\micro\meter} with a mean radius of \SI{6.98}{\micro\meter}.
This setup and the model for blood vessel collapse in the discrete case are taken from Kremheller. \cite[pp157--162]{Kremheller2021Thesis}
The tumour, as shown in Figure~\ref{fig:WithArteries}A, exhibits the same characteristics as the idealised spherical tumour, in particular, a non-perfused core with collapsed blood vessels.
We employ the same treatment protocol as in the previous example, with the same parameters for the nanoparticle transport and hyperthermia treatment (see Table~\ref{Tab:ParametersNanoparticleHyperthermia}).
The mass fraction of injected nanoparticles is $\omegaNPvD = \num{2.0e-3}$, and the specific absorption rate is $SAR = \SI{2.0}{\mega\watt\per\kilo\gram}$.
The initial temperature is again set to $T_{b} = \SI{37}{\celsius}$.

As illustrated in Figure~\ref{fig:WithArteries}B, the mass fraction of nanoparticles in the IF varies across different time points.
Initially, nanoparticles accumulate around the blood vessels, subsequently diffusing into the interstitial spaces between them.
However, the nanoparticles do not reach a significant proportion of the tumour mass:
because of the collapsed blood vessels in the tumour core, substantially fewer nanoparticles reach this area.
Finally, after the injection phase, the nanoparticles are drained by the lymphatic system, leading to a decrease in the mass fraction.

The temperature field is shown in Figure~\ref{fig:WithArteries}C:
the temperature increases during the nanoparticle injection phase and reaches a maximum of $\SI{41.2}{\celsius}$ at \SI{60}{\minute}.
As expected, the highest temperature is reached where the nanoparticles accumulate, i.e., around the blood vessels.
At the right boundary of the domain, where no nanoparticles are located, the temperature is significantly lower.
This demonstrates the challenge of applying nanoparticle-mediated hyperthermia treatment to a tumour:
the nanoparticles do not reach the tumour core, and the temperature increase is limited to the area around the blood vessels.
Hence, the tumour core is not heated sufficiently, while the surrounding healthy tissue is exposed to high and potentially damaging temperatures.

To further investigate the influence of the heat sink due to blood perfusion, we compare the temperature field to the results neglecting the cooling effect of blood perfusion.
Visually, the temperature field is identical for both cases and therefore not shown.
While the original studies \cite{Nabil2015,Nabil2016} claim the superiority of the discrete model over the lumped form of the Pennes' bioheat equation, our results show that the heat exchange coefficient used in these studies together with a realistic microvascular network does not result in a significant change in temperature compared to neglecting the cooling effect of blood perfusion.
Note that we not only use the same value for the heat exchange coefficient but that the radii of the blood vessels are also in the same range as in the original studies.
Increasing the heat exchange coefficient by two orders of magnitude, to a hypothetical values of $\beta_T = \SI{2e-3}{\watt\per\milli\meter\squared\per\kelvin}$, results in a significant change, as shown in Figure~\ref{fig:WithArteries}D:
the temperature is reduced by up to \SI{2.5}{\celsius} compared to the results in Figure~\ref{fig:WithArteries}C.
Increasing the heat exchange coefficient by two orders of magnitude is equivalent to the cooling effect of blood vessels with a mean radius of \SI{0.7}{\milli\meter}, i.e., larger arteries, which are not part of the microvasculature. \cite{Muller2008}

Finally, we compare the cooling effect of the discrete model to the lumped heat sink term based on Pennes' bioheat equation.
Figure~\ref{fig:WithArteries}E shows the temperature curves at $y = \SI{1.8}{\milli\meter}$ after \SI{60}{\minute} for different parameters of the heat exchange coefficient or the blood perfusion rate.
We also include the case without the cooling effect of blood perfusion.
The discrete model with a heat exchange coefficient of $\beta_T = \SI{2e-5}{\watt\per\milli\meter\squared\per\kelvin}$, as proposed in the literature, \cite{Nabil2015,Nabil2016} predicts a maximum temperature decrease of \SI{0.05}{\celsius} compared to the case without blood perfusion.
In contrast, the lumped model based on Pennes' bioheat equation predicts a maximum temperature decrease of \SI{0.4}{\celsius} or \SI{0.75}{\celsius} for a blood perfusion rate of $w = \SI[per-mode=reciprocal]{0.018}{\per\second}$ or $w = \SI[per-mode=reciprocal]{0.036}{\per\second}$, respectively.
Hence, the lumped model overestimates the cooling effect of blood perfusion by a factor of 8 to 15 for parameters typically used in the literature.

These results lead to the following conclusion:
small capillaries do not have a significant cooling effect during nanoparticle-mediated hyperthermia treatment, which confirms the results of previous studies in the literature. \cite{Weinbaum1985,Yuan2008}
Only larger vessels, which are not present in the tumour microenvironment, have a significant impact on the temperature.
For a microvascular network, the lumped heat sink term in Pennes' bioheat equation significantly overestimates the heat exchange due to blood perfusion.
However, we only studied one specific configuration of a microvascular network of a rat experimentally measured in two dimensions.
Investigating different \textit{in vivo} configurations in three dimensions is clearly necessary to draw a more general conclusion.
Our model of the tumour and its vasculature provide a perfect framework to do so, as shown in our previous study. \cite{Kremheller2021}

\subsection{\textit{In vivo} tumour in a mouse model}
\label{sec:MouseModel}

In the final example, we investigate nanoparticle-mediated hyperthermia treatment in an \textit{in vivo} tumour in a mouse model.
This example is based on the experimental study by Cho et al. \cite{Cho2017} where the authors investigated hyperthermia treatment with assemblies of iron oxide nanocubes (ANC) in a mouse bearing glioblastoma cells.
The results in the original publication \cite{Cho2017} show that the ANCs accumulate in clusters in the tumour.
Our aim therefore is to investigate how the accumulation in clusters influences the temperature during hyperthermia treatment.

\begin{figure}[btp]
    \centering
    \includegraphics[width=\textwidth]{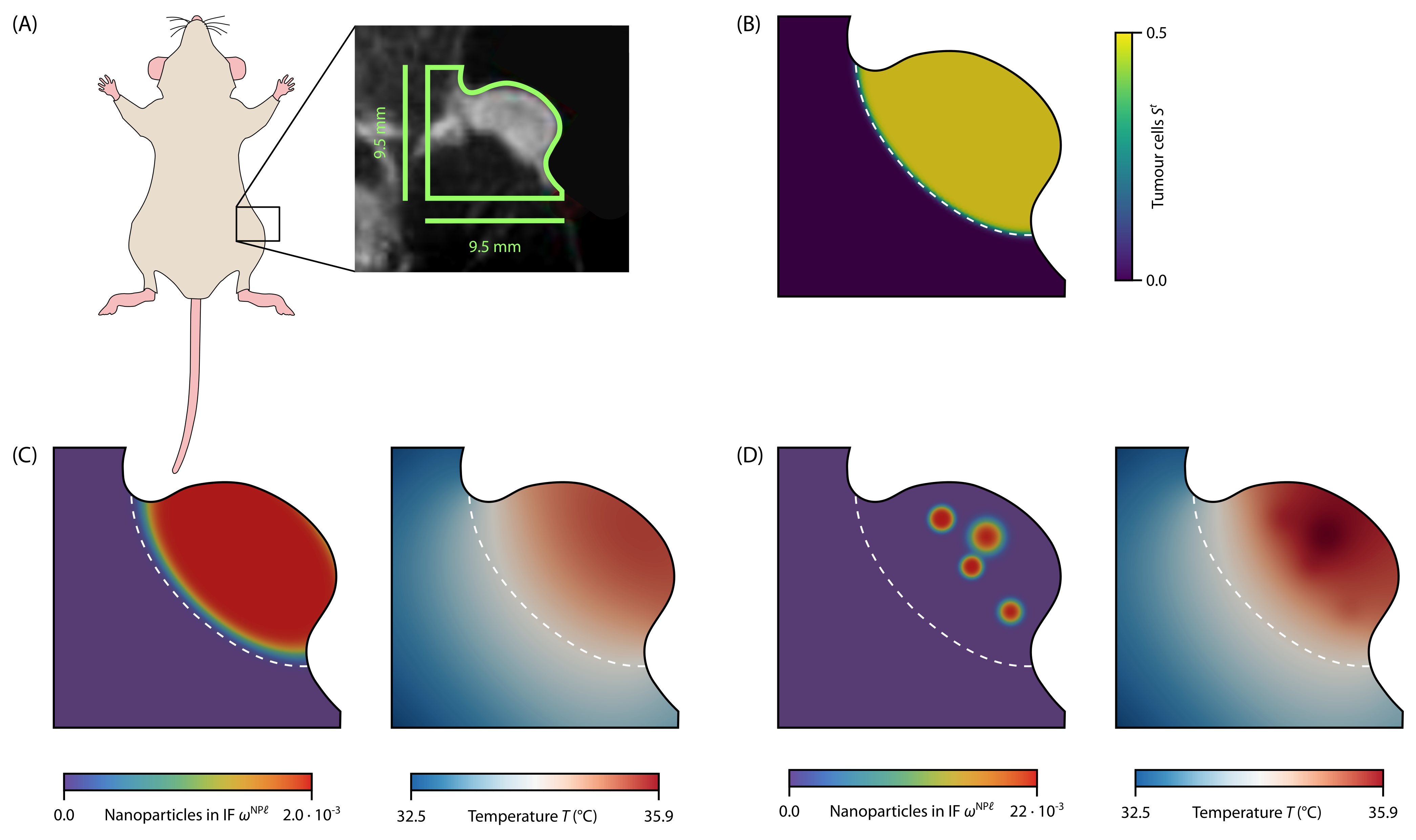}
    \caption{In vivo tumour in a mouse model.
        (A)~Geometry of the tumour in the leg of the mouse taken from the experimental study of Cho et al. \cite{Cho2017}
        (B)~Saturation of tumour cells $S^t$.
        The white dashed line indicates the tumour boundary in all plots.
        (C)~Homogeneous distribution: Mass fraction of nanoparticles in the interstitial fluid (IF) $\omegaNPlD$ and temperature field after \SI{30}{\minute}.
        (D)~Clustered distribution: Mass fraction of nanoparticles in the IF $\omegaNPlD$ and temperature field after \SI{30}{\minute}.
    }
    \label{fig:Mouse}
\end{figure}

We segment the geometry of the tumour in the leg of the mouse from the magnetic resonance image, as shown in Figure~\ref{fig:Mouse}A.
The domain has a size of $\SI{9.5}{\milli\meter} \times \SI{9.5}{\milli\meter}$ with a tumour size of \SI{8}{\milli\meter} along the major axis and \SI{4}{\milli\meter} along the minor axis.
We discretise the domain with 163840 elements.
As we are only interested in the temperature in this example, we prescribe the saturation of tumour cells and the mass fraction of nanoparticles in the IF as Dirichlet boundary conditions on the entire domain.
The prescribed saturation of tumour cells is shown in Figure~\ref{fig:Mouse}B.
Since the previous example showed that the tumour microvasculature does not have a significant cooling effect on the temperature, we neglect the heat sink due to blood perfusion in this example.
Thus, the heat exchange coefficient $\beta_T$ refers only to the heat exchange with the surrounding tissue, as included in the Robin-type boundary condition given by Equation~(\ref{Eq:RobinBoundaryCondition}).
For the heat exchange with air at the outer boundary of the domain, we assume a lower heat exchange coefficient.
For the thermal properties of the tissue, we use the same values as in the previous examples.
The specific absorption rate is $SAR = \SI{0.1098}{\mega\watt\per\kilo\gram}$ and the initial temperature $T_{b} = \SI{29}{\celsius}$, as measured in the experiment. \cite{Cho2017}
Table~\ref{Tab:ParametersNanoparticleHyperthermiaMouse} lists the parameters we use in this example.

\begin{table}[btp]
    \caption{Parameters for hyperthermia treatment in the \textit{in vivo} tumour in a mouse model (Section~\ref{sec:MouseModel}).}
    \label{Tab:ParametersNanoparticleHyperthermiaMouse}
    \centering
    \begin{tabular}{llllll}
        \toprule
        \multicolumn{2}{l}{Parameter}    & Value                         & Unit                  & Source                                                                                         \\
        \midrule
        $T_{b}$                          & Body temperature of the mouse & $\num{29}$            & \si{\celsius}                                                  & \citenum{Cho2017}             \\
        $c_p^\gamma \;\forall \gamma$    & Tissue-specific heat capacity & $\num{3470}$          & \si[per-mode=symbol]{\joule\per\kilo\gram\per\kelvin}          & \citenum{Nabil2015,Huang2010} \\
        $\kappa^\gamma \;\forall \gamma$ & Tissue thermal conductivity   & $\num{0.51e-3}$       & \si[per-mode=symbol]{\watt\per\milli\meter\per\kelvin}         & \citenum{Cervadoro2013}       \\
        $SAR$                            & Specific absorption rate      & $\num{0.1098}$        & \si[per-mode=symbol]{\mega\watt\per\kilo\gram}                 & \citenum{Cho2017}             \\
        $\beta_T$                        & Heat exchange coefficient     & $\num{2e-5}$ (tissue) & \si[per-mode=symbol]{\watt\per\milli\meter\squared\per\kelvin} & \citenum{Nabil2016}           \\
                                         &                               & $\num{0.3e-5}$ (air)  & \si[per-mode=symbol]{\watt\per\milli\meter\squared\per\kelvin} & Assumption                    \\
        \bottomrule
    \end{tabular}
\end{table}

To investigate the influence of the clustering of nanoparticles on the temperature, we compare the results for a homogeneous distribution of nanoparticles in the IF to a clustered distribution.
In the homogeneous case, we prescribe a constant mass fraction of nanoparticles in the IF of $\omegaNPlD = \num{2.0e-3}$ in the tumour area, as shown in Figure~\ref{fig:Mouse}C.
In the clustered case, we prescribe a locally higher mass fraction of nanoparticles such that the nanoparticles are accumulated in four clusters in the tumour area, as shown in Figure~\ref{fig:Mouse}D.
Since the results for the idealised spherical tumour showed that the amount of nanoparticles significantly influences the temperature, we set the clustered distribution such that the integrated mass of nanoparticles in the tumour area is the same as in the homogeneous case.
We analyse the temperature after \SI{30}{\minute} similar to the experiments.

The results show that the temperature increases from \SI{29}{\celsius} to almost \SI{36}{\celsius} in both cases, i.e., a temperature increase of \SI{7}{\celsius}.
The temperature in the clustered case is locally \SI{0.5}{\celsius} higher than in the homogeneous case.
These results are similar to the experimental results of Cho et al., \cite{Cho2017} who measured a temperature increase of \SI{5}{\celsius} at the surface of the tumour and expected a further increment of \SI{3}{\celsius} in the tumour core.
Note that the mass fraction of nanoparticles in the tumour is not known in the experiment.
Only the mass of intravenously injected ANCs per kilogram body weight is given together with the coarse magnetic resonance imaging data.
Experimentally, the exact temperature field in the tumour cannot be measured, so the influence of nanoparticle clustering cannot be studied.
Here, simulations can give valuable insights that experimental studies alone cannot provide.

\section{Conclusions}\label{sec4}

In this paper, we presented a computational model for the simulation of nanoparticle-mediated hyperthermia treatment of tumours.
Our model is based on and fully integrated with a multiphase porous-media model of the tumour and its microenvironment.
We considered nanoparticle transport in the tumour microenvironment and hyperthermia treatment, with particular emphasis on modelling the cooling effect of blood perfusion.
Our results showed that the temperature reached in the tumour highly depends on the amount of nanoparticles accumulated in the tumour and the specific absorption rate of the nanoparticles.
Further, host tissue surrounding the tumour is also exposed to considerable doses of heat due to the high thermal conductivity of the tissue, causing pain or even irreversible damage.
Using a discrete model of a realistic microvasculature, we found that small capillaries do not have a significant cooling effect on the tumour temperature and that the lumped heat sink model in Pennes' bioheat equation, with values typically used in the literature, significantly overestimates heat exchange due to blood perfusion.
Finally, we showed that the clustering of nanoparticles in the tumour can lead to a slightly higher temperature than a homogeneous distribution.
However, the overall temperature is similar if the total mass of nanoparticles in the tumour area is the same.

To successfully apply hyperthermia treatment in cancer therapy, it is crucial to precisely control the temperature in the tumour to avoid under- or overtreatment.
While we focused here on nanoparticle-mediated hyperthermia treatment, our model is equally applicable to other cases of local hyperthermia, such as the external, interluminal, or endocavitary and interstitial approaches.
Computational models can predict the temperature field in the tumour and surrounding tissue, provide insights not accessible by experiments, and help to optimise the treatment.

%\backmatter
\bmsection*{Author contributions}

All authors contributed to the study conceptualisation and design.
BW performed the development of methodology, implementation, formal analysis, visualisation, and wrote the original draft of the manuscript.
All authors commented on and critically reviewed previous versions of the manuscript.
All authors read and approved the final manuscript.

\bmsection*{Funding statement}

BAS was supported by the Visiting Fellowship for Alumni Fellows of the Institute for Advanced Study (IAS).
WAW was supported by BREATHE, a Horizon 2020 | ERC-2020-ADG project, grant agreement No. 101021526.

\bmsection*{Data availability}

The code used for the simulations in this study is implemented in the open-source software package 4C, which is available at \href{https://github.com/4C-multiphysics/4C}{github.com/4C-multiphysics/4C}.
All other data is available from the corresponding author upon reasonable request.

\bmsection*{Ethics approval}

Not applicable.

\bmsection*{Conflict of interest}

The authors declare no potential conflict of interests.

% \bmsection*{Supporting information}

% Additional supporting information may be found in the
% online version of the article at the publisher’s website.

\appendix

\bmsection{Supplementary derivation} \label{App:SuppEquations}

We here provide the detailed derivation of Equations (\ref{Eq:HeatEquationTransformed}).
The energy balance is given by
\begin{equation}
    c_{p}^{\gamma} \frac{\partial \left( \rho^\gamma \varepsilon^\gamma T \right)}{\partial t}
    \bigg\arrowvert_{\vct{x}}
    +
    c_{p}^{\gamma} \vct{\nabla} \cdot \left( \rho^\gamma \varepsilon^\gamma \, T \vct{v}^\gamma \right)
    -
    \vct{\nabla} \cdot \left( \kappa^\gamma \varepsilon^\gamma \vct{\nabla} T \right)
    =
    \varepsilon^\gamma \left(Q_p - Q_{bl}\right).
\end{equation}
Applying the product rule to the time derivative term and to the convergence term, we obtain
\begin{equation}
    c_{p}^{\gamma} \rho^\gamma \varepsilon^\gamma \frac{\partial T}{\partial t}
    \bigg\arrowvert_{\vct{x}}
    +
    c_{p}^{\gamma} T \frac{\partial \left( \rho^\gamma \varepsilon^\gamma \right)}{\partial t}
    \bigg\arrowvert_{\vct{x}}
    +
    c_{p}^{\gamma} T \, \vct{\nabla} \cdot \left( \rho^\gamma \varepsilon^\gamma \vct{v}^\gamma \right)
    +
    c_{p}^{\gamma} \rho^\gamma \varepsilon^\gamma \vct{v}^\gamma \cdot \vct{\nabla} T
    -
    \vct{\nabla} \cdot \left( \kappa^\gamma \varepsilon^\gamma \vct{\nabla} T \right)
    =
    \varepsilon^\gamma \left(Q_p - Q_{bl}\right),
    \label{Eq:EnergyBalanceProductRule}
\end{equation}
where we assume the phases to be incompressible, i.e., $\rho^\gamma = \rho^\gamma_0 = \text{const}$ and, hence, $\partial \rho^\gamma /\partial t = 0$.
The spatial time derivative of a spatial quantity $g$ can be related to the material time derivative by
\begin{equation}
    \frac{\partial g}{\partial t} \bigg\arrowvert_{\vct{X}}
    = \frac{\partial g}{\partial t} \bigg\arrowvert_{\vct{x}} + \vct{v}^\mathrm{s} \cdot  \vct{\nabla} g,
    \label{Eq:SpatialTimeDerivative}
\end{equation}
where the velocity of the solid phase is given by
\begin{equation}
    \vct{v}^\mathrm{s}
    = \frac{\partial \vct{x}}{\partial t} \bigg\arrowvert_{\vct{X}}
\end{equation}
and can be interpreted as the velocity of the spatial configuration as it is seen from the material configuration of the solid phase.
This allows rewriting the first term in Equation~(\ref{Eq:EnergyBalanceProductRule}) and gives
\begin{equation}
    c_{p}^{\gamma} \rho^\gamma \varepsilon^\gamma \frac{\partial T}{\partial t}
    \bigg\arrowvert_{\vct{X}}
    -
    c_{p}^{\gamma} \rho^\gamma \varepsilon^\gamma \vct{v}^\mathrm{s} \cdot  \vct{\nabla} T
    +
    c_{p}^{\gamma} \rho^\gamma \varepsilon^\gamma \vct{v}^\gamma \cdot \vct{\nabla} T
    -
    \vct{\nabla} \cdot \left( \kappa^\gamma \varepsilon^\gamma \vct{\nabla} T \right)
    =
    \varepsilon^\gamma \left(Q_p - Q_{bl}\right)
    -
    c_{p}^{\gamma} T
    \left[
        \frac{\partial \left( \rho^\gamma \varepsilon^\gamma \right)}{\partial t}
        \bigg\arrowvert_{\vct{x}}
        +
        \vct{\nabla} \cdot \left( \rho^\gamma \varepsilon^\gamma \vct{v}^\gamma \right)
        \right]
    ,
    \label{Eq:EnergyBalanceSpatialTimeDerivative}
\end{equation}
where we brought the second and the third term of Equation~(\ref{Eq:EnergyBalanceProductRule}) to the left-hand side of the equation.
Finally, the mass balance equation for an arbitrary phase $\gamma$ on the macroscale based on TCAT is written as
\begin{equation}
    \frac{\partial \left(\varepsilon^\gamma \rho^\gamma \right)}{\partial t} \bigg\arrowvert_{\vct{x}}
    + \vct{\nabla} \cdot \left( \rho^\gamma \varepsilon^\gamma \vct{v}^\gamma \right)
    =
    \sum\limits_{\kappa \in \mathcal{J}_{c\gamma}} \overset{\kappa \rightarrow \gamma}{M}
    \label{Eq:MassBalancePhase}
\end{equation}
with $\varepsilon^\gamma$ being the volume fraction, $\rho^\gamma$ the density and $\vct{v}^\gamma$ the velocity of phase $\gamma$.
The mass transfer term on the right-hand side of the equation denotes the mass exchange terms representing transport of mass at the interface $\mathcal{J}$ between the phases $\kappa$ and $\gamma$.
We can therefore rewrite Equation~(\ref{Eq:EnergyBalanceSpatialTimeDerivative}) as
\begin{equation}
    c_{p}^{\gamma} \rho^\gamma \varepsilon^\gamma \frac{\partial T}{\partial t}
    \bigg\arrowvert_{\vct{X}}
    +
    c_{p}^{\gamma} \rho^\gamma \varepsilon^\gamma \left( \vct{v}^\gamma - \vct{v}^\mathrm{s} \right) \cdot \vct{\nabla} T
    -
    \vct{\nabla} \cdot \left( \kappa^\gamma \varepsilon^\gamma \vct{\nabla} T \right)
    =
    \varepsilon^\gamma \left(Q_p - Q_{bl} \right)
    -
    c_{p}^{\gamma} T
    \sum\limits_{\kappa \in \mathcal{J}_{c\gamma}} \overset{\kappa \rightarrow \gamma}{M}
    .
    \label{Eq:EnergyBalanceSpatialMassBalance}
\end{equation}
For the solid phase, the convective term cancels out, i.e.,
\begin{equation}
    c_{p}^s \rho^s \varepsilon^s \frac{\partial T}{\partial t}
    \bigg\arrowvert_{\vct{X}}
    -
    \vct{\nabla} \cdot \left( \kappa^s \varepsilon^s \vct{\nabla} T \right)
    =
    \varepsilon^s \left(Q_p - Q_{bl} \right)
    -
    c_{p}^s T
    \sum\limits_{\kappa \in \mathcal{J}_{cs}} \overset{\kappa \rightarrow s}{M}
    .
\end{equation}

\end{document}